\documentclass[twocolumn,aps,floatfix,final,prl,showpacs,superscriptaddress]{revtex4}
\usepackage[latin9]{inputenc}
\usepackage{amsmath}
\usepackage{amssymb}
\usepackage{graphicx}

\makeatletter

\providecommand{\tabularnewline}{\\}

\@ifundefined{textcolor}{}
{%
 \definecolor{BLACK}{gray}{0}
 \definecolor{WHITE}{gray}{1}
 \definecolor{RED}{rgb}{1,0,0}
 \definecolor{GREEN}{rgb}{0,1,0}
 \definecolor{BLUE}{rgb}{0,0,1}
 \definecolor{CYAN}{cmyk}{1,0,0,0}
 \definecolor{MAGENTA}{cmyk}{0,1,0,0}
 \definecolor{YELLOW}{cmyk}{0,0,1,0}
}

\usepackage{epsfig}

\makeatother

\begin{document}

\title{Quantum Oscillations in EuFe$_{2}$As$_{2}$ single crystals}

\author{P. F. S. Rosa}

\affiliation{Instituto de F\'{i}sica \lq\lq Gleb Wataghin\rq\rq, UNICAMP,
Campinas-SP, 13083-859, Brazil.}
\affiliation{University of California, Irvine, California 92697-4574, USA.}

\author{B. Zeng}

\affiliation{National High Magnetic Field Laboratory, Florida State University, Tallahassee, Florida 32310, USA.}

\author{C. Adriano}

\affiliation{Instituto de F\'{i}sica \lq\lq Gleb Wataghin\rq\rq, UNICAMP,
Campinas-SP, 13083-859, Brazil.}

\author{T. M. Garitezi}

\affiliation{Instituto de F\'{i}sica \lq\lq Gleb Wataghin\rq\rq, UNICAMP,
Campinas-SP, 13083-859, Brazil.}

\author{T. Grant}

\affiliation{University of California, Irvine, California 92697-4574, USA.}

\author{Z. Fisk}

\affiliation{University of California, Irvine, California 92697-4574, USA.}

\author{L. Balicas}

\affiliation{National High Magnetic Field Laboratory, Florida State University, Tallahassee, Florida 32310, USA.}

\author{R. R. Urbano}

\affiliation{Instituto de F\'{i}sica \lq\lq Gleb Wataghin\rq\rq, UNICAMP,
Campinas-SP, 13083-859, Brazil.}

\author{P. G. Pagliuso}

\affiliation{Instituto de F\'{i}sica \lq\lq Gleb Wataghin\rq\rq, UNICAMP,
Campinas-SP, 13083-859, Brazil.}

\date{\today}
\begin{abstract}
Quantum oscillation measurements can provide important information about the Fermi surface (FS) properties of strongly correlated metals. Here,  we report a Shubnikov-de Haas (SdH) effect study on the pnictide parent compounds EuFe$_{2}$As$_{2}$ (Eu122) and BaFe$_{2}$As$_{2}$ (Ba122) grown by In-flux. Although both members are isovalent compounds with approximately the same density of states at the Fermi level, our results reveal subtle changes in their fermiology. Eu122 displays a complex pattern in the Fourier spectrum, with band splitting, magnetic breakdown orbits, and effective masses sistematically larger when compared to Ba122, indicating that the former is a more correlated metal. Moreover, the observed pockets in Eu122 are more isotropic and 3D-like, suggesting an equal contribution from the Fe $3d$ orbitals to the FS. We speculate that these FS changes may be responsible for the higher spin-density wave ordering temperature in  Eu122.
\end{abstract}


\maketitle

\section{\label{sec:intro}Introduction}

The Fe-based superconductors (SC) are a subject of intensive investigations owing to their high SC transition temperature ($T_{c}$) up to $56$ K \cite{Kamihara,Rotter,K,Gd}. Although they all present a common structural element, the FeAs layers, there is a variety of dissimilarities in their physical properties, such as magnetic ordered moments, effective masses, size of SC gaps and the $T_{c}$
 itself \cite{Review}. For instance, 
KFe$_{2}$As$_{2}$
is a non-magnetic metal which superconducts below $4$ K while BaFe$_{2}$As$_{2}$ (Ba122)
has a transition to a spin density wave (SDW) metallic state at $139$ K, closely related to a structural
transition \cite{Urbano}. In the latter, SC can be induced by hydrostatic pressure and/or chemical substitution (e.g. K, Co, Ni, Cu, and Ru). A deep understanding of such differences is still a central open question that remains to be answered. In this regard, it is essential to unveil the Fermi surface (FS) geometry and symmetry, which in turn affect the nature of the magnetic fluctuations in momentum space. Complementarily, it is desirable to unravel the role
of local distortions in determining these FS properties.

Experimentally, the most widely employed technique to directly access the electronic structure of crystals is angle resolved photoemission spectroscopy
(ARPES). For undoped 122-parent compounds $A$Fe$_{2}$As$_{2}$ ($A =$ Ba, Ca, Eu, Sr, ...), ARPES data usually display five Fe $3d$ FS sheets in the paramagnetic regime: two electron-like bands around the $M$-point and three hole-like bands around the $\Gamma$-point of the Brillouin zone (BZ). Furthermore, polarized ARPES measurements show the predominance of $t_{2g}$ orbitals ($xy$, $xz/yz$) at the FS with the outer hole-pocket being mostly $xy$, in agreement with DFT + DMFT calculations \cite{Zhou_ARPES, Jong_ARPES, Ding, KH}. Interestingly, a comparison between Ba122 and Eu122 in the paramagnetic state reveals that the outer hole $xy$ pocket is two times larger in Eu122 as
compared to Ba122 \cite{Setti}, suggesting an increase of the planar $xy$ occupation in the latter.  It is noteworthy that, although the total Fe occupation remains the same, it is possible to change the occupation between distinct $t2g$, or even $eg$, orbitals. Such symmetry changes are directly carried to the FS and their influence is not fully understood yet. In addition, as the temperature is lowered below $T_{SDW}$, further striking consequences appear for the FS, such as reconstruction, band splitting, opening of gaps, etc.
  However, ARPES is a surface sensitive technique and it is highly desirable to confront its data with bulk FS experiments since the surface electronic structure may differ from the bulk one. 

In this context, a precise FS determination by a bulk-sentivite technique become crucial and quantum oscillations (QO) measurements have been proven to be an important complementary technique to ARPES \cite{Book,Sebastian1}. 
In particular, low temperature QO experiments
have been reported on several members of the 122 family 
$A$Fe$_{2}$As$_{2}$ ($A =$ Ba, Ca, Sr) providing high momentum space resolution at the FS and important information on the  effective masses \cite{Sebastian2,Analytis,Terashima,Harrison}. In particular, five low frequency Fourier peaks
have been observed for detwinned BaFe$_{2}$As$_{2}$ at
$F_{\delta}=500$ T, $F_{\alpha}=440$ T, $F_{\beta}=170$ T, and
$F_{\gamma}=90$ T, each occupying $~ 2$ \% of nonmagnetic BZ. However, up to date,
no QO data have been reported for the Eu-member although single crystals with a residual resistivity
ratio (RRR) as high as $15$ have been synthesized \cite{Kurita}.
 The semi-metal compound EuFe$_{2}$As$_{2}$ (Eu122) is a unique member displaying a SDW phase transition at $190$ K and also an antiferromagnetic phase transition due to the Eu$^{2+}$ ions at $T_{N}=19$ K. Interestingly, Eu122
presents a metamagnetic anomaly at $H_{m} \sim 2$ T, corresponding to the transition from a Eu$^{2+}$ AFM state to a FM one. Thus, there is no additional reconstruction due to the Eu$^{2+}$ ordering at low temperatures.

In this paper, we report a comparative Shubnikov de-Haas (SdH) study between Eu122 and Ba122 single crystals grown by In-flux. Our results for the Ba-member are in agreement with previous reports 
\cite{Analytis,Sebastian2} and the unprecedented observation of QO in the Eu-member indicates the synthesis of high quality crystals by the In-flux technique. Our results reveal subtle changes in the fermiology of Eu122 when compared to Ba122, including band splitting, enhancement of effective masses and a significant increase in the isotropic and three-dimensional character of the bands.
Our findings shed new light on the understanding of the SDW state and provide a scenario for the
observed differences in the physical properties of this class of materials.

\section{\label{sec:exp} Experimental Details}

Single-crystalline samples of EuFe$_{2}$As$_{2}$ were grown using the flux technique with starting composition Eu:Fe:As:In=1:2:2:25. The mixture was placed in an alumina crucible and sealed in a quartz tube under vacuum. The sealed tube was heated up to $1100 ^{\circ}\mathrm{C}$ for 8 h and then cooled down to $1000 ^{\circ}\mathrm{C}$ at  $2 ^{\circ}\mathrm{C/h}$. 
The furnace was then shutted off and the excess of In flux was removed at $400 ^{\circ}\mathrm{C}$ 
by centrifugation. Single-crystalline samples of BaFe$_{2}$As$_{2}$ were grown as described in Ref. \cite{Garitezi}. 
Our crystals were checked by x-ray powder diffraction and 
elemental analysis using a commercial Energy Dispersive Spectroscopy
(EDS) microprobe. Specific-heat measurements were performed in a Quantum Design physical properties measurement system (PPMS) small-mass calorimeter that employs a quasiadiabatic thermal relaxation technique.The in-plane
resistivity ($\rho_{ab}$ ($T$)) was measured using a standard four-probe
method inside a $3$He cryostat. Magnetic fields up to $35$ T and $45$ T  were applied at the National High Magnetic Field Laboratory, Tallahassee. The angle between the sample and the external field was measured with a Hall probe.

\section{\label{sec:results} Results and Discussion}

Fig. 1 displays $\rho_{ab}$ ($T$) at zero magnetic field for Eu122 and Ba122 single
crystals. A linear metallic behavior is observed down to $T_{SDW}$ where a sudden drop can be identified in the curves at $193$ K and $139$ K, respectively. The transition to a SDW state can be also seen as a sharp peak in the specific heat of both compounds (right inset of Fig. 1).
 In addition, as the temperature is further decreased, a second drop is observed in Eu122 due to the Eu$^{2+}$ antiferromagnetic
ordering at $19$ K (left inset of Fig. 1). The residual resistivity values and RRR are $10$ $\mu\Omega$.cm and $25$ for Eu122, respectively, indicating a high degree of crystallinity.  It is noteworthy that the obtained transition temperature for Eu122 is slightly higher than the one observed for single crystals grown by Sn- and FeAs-fluxes ($T_{SDW}=189$ K) \cite{Jeevan,SF}. Interestingly, higher $T_{SDW}$ and/or $T_{c}$ is
a general trend obtained in our In-grown single crystals \cite{Garitezi}.

\begin{figure}[!ht]
\includegraphics[width=0.4\textwidth, bb=0 0 800 600]{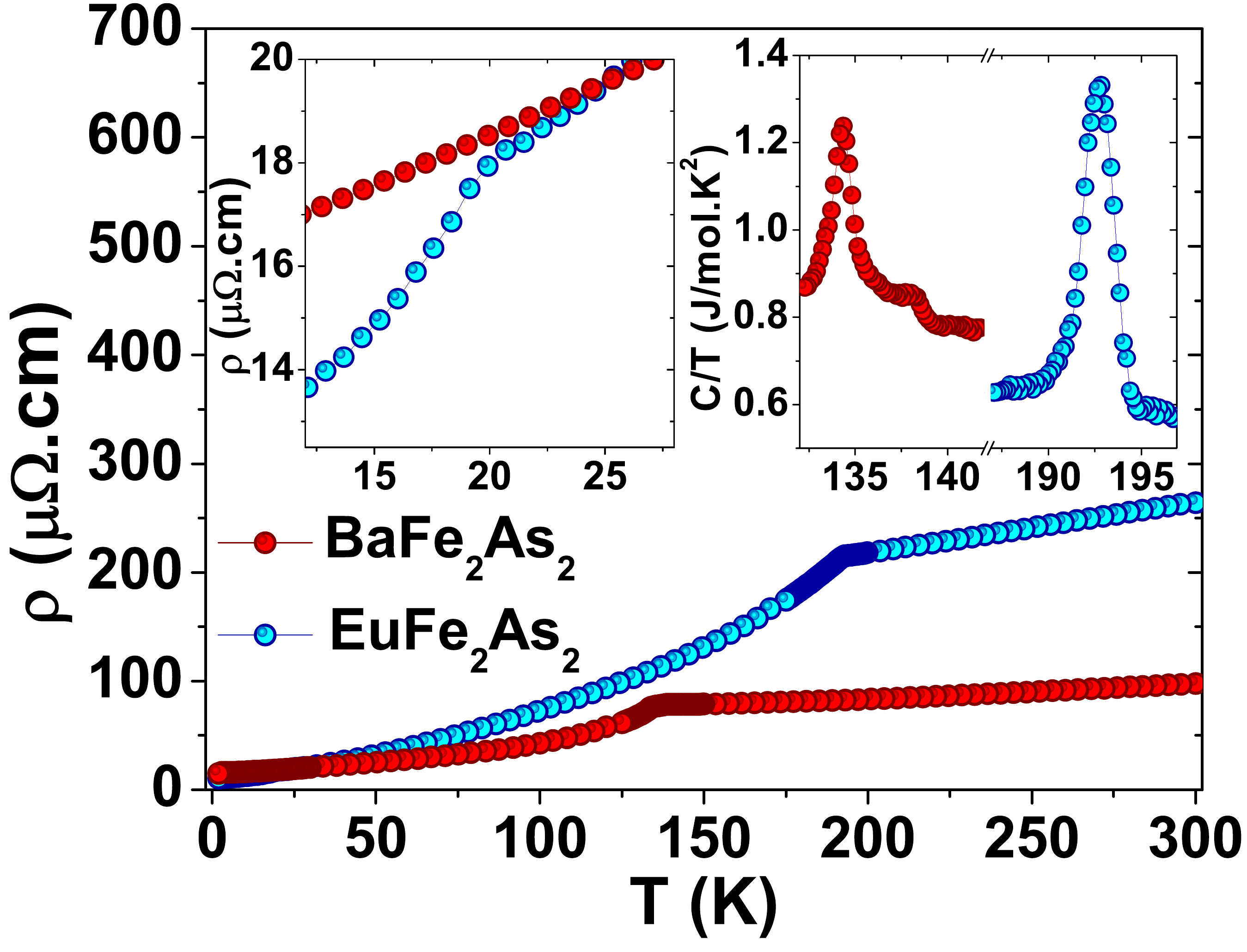} 
\caption{$\rho_{ab}$ (T) for Eu122 and Ba122 single crystals. The left inset shows the Eu$^{2+}$ AFM ordering temperature and the right inset shows the SDW anomaly in the specific heat.}
\label{Fig1} 
\end{figure}

Figs. 2a and 2b show $\rho_{ab}$ (T) at $0.3$ K as a function of the applied magnetic field for Eu122 and Ba122 single crystals, respectively. The background was subtracted using a smooth 4th order polynomial and, as a result, anisotropic QO are clearly observed above $\sim 15$ T.  Fig. 2c and 2d display the Fast Fourier Transform (FFT) of the above oscillations as a function of frequency in Eu122 and Ba122 single crystals, respectively. Five low frequency peaks can be identified for both compounds and are listed in Table I. The labeling of these branches is identical to Ref. \cite{Sebastian1}.

\begin{figure}
 \begin{tabular}{rr}
 \begin{minipage}{0.5\textwidth}
\hspace*{-5.15cm}
\includegraphics[width=0.5\textwidth,bb=0 0 800 600]{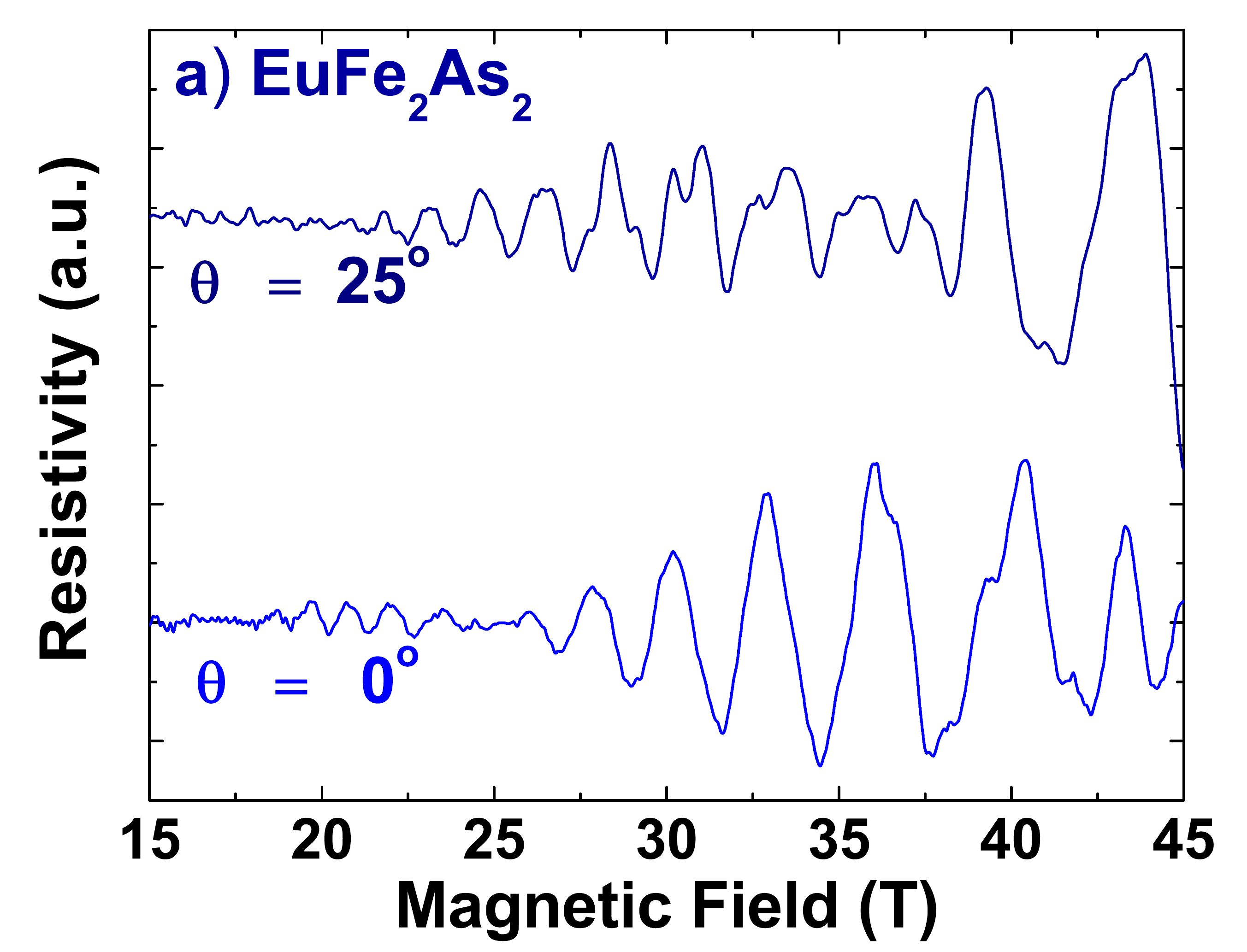} 
 \end{minipage}&
 \begin{minipage}{0.5\textwidth}
\hspace*{-14.5cm}
 \includegraphics[width=0.5\textwidth,bb=0 0 800 600]{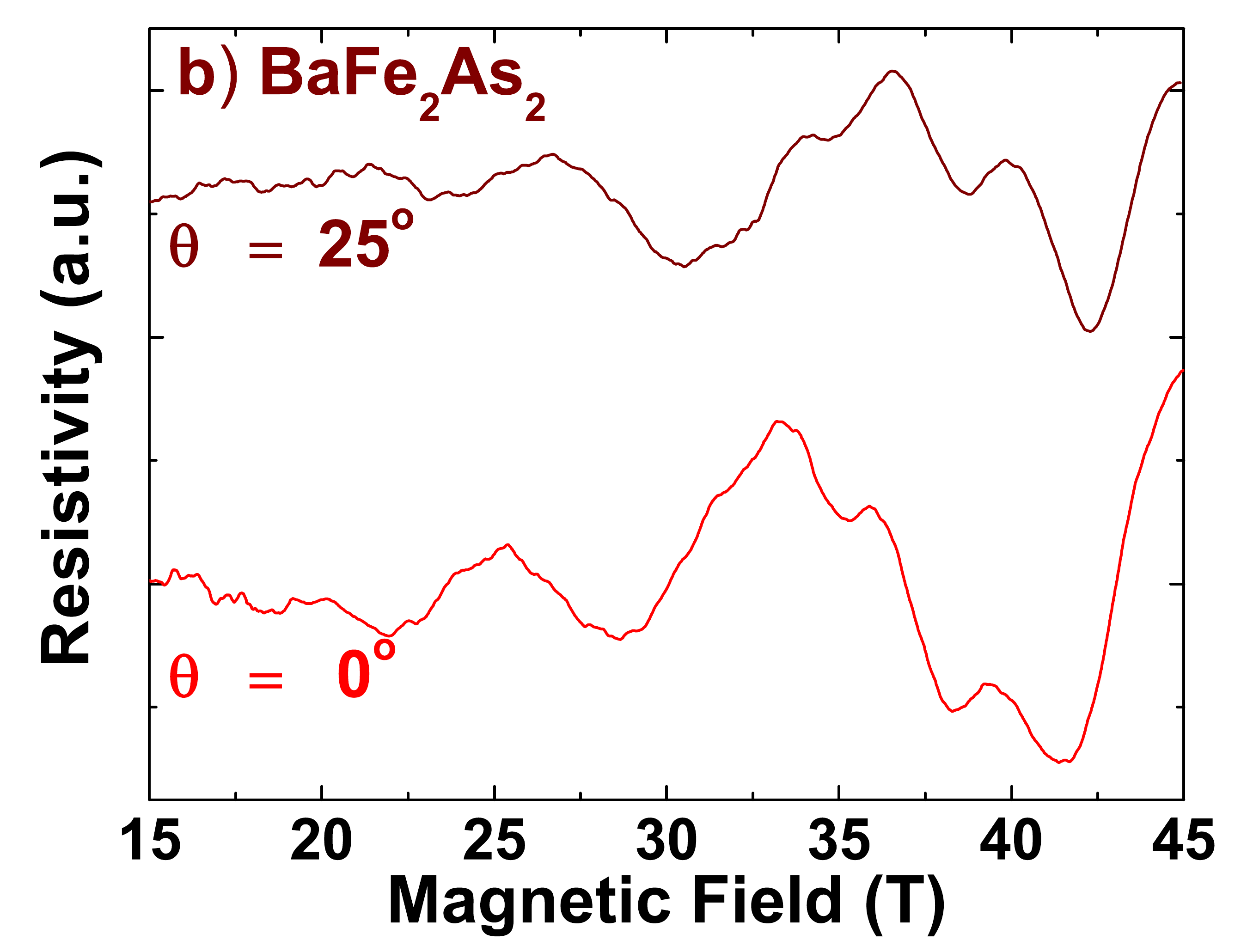} 
\end{minipage}\\
 \begin{minipage}{0.5\textwidth}
\hspace*{-5.05cm}
\includegraphics[width=0.5\textwidth,bb=0 0 800 600]{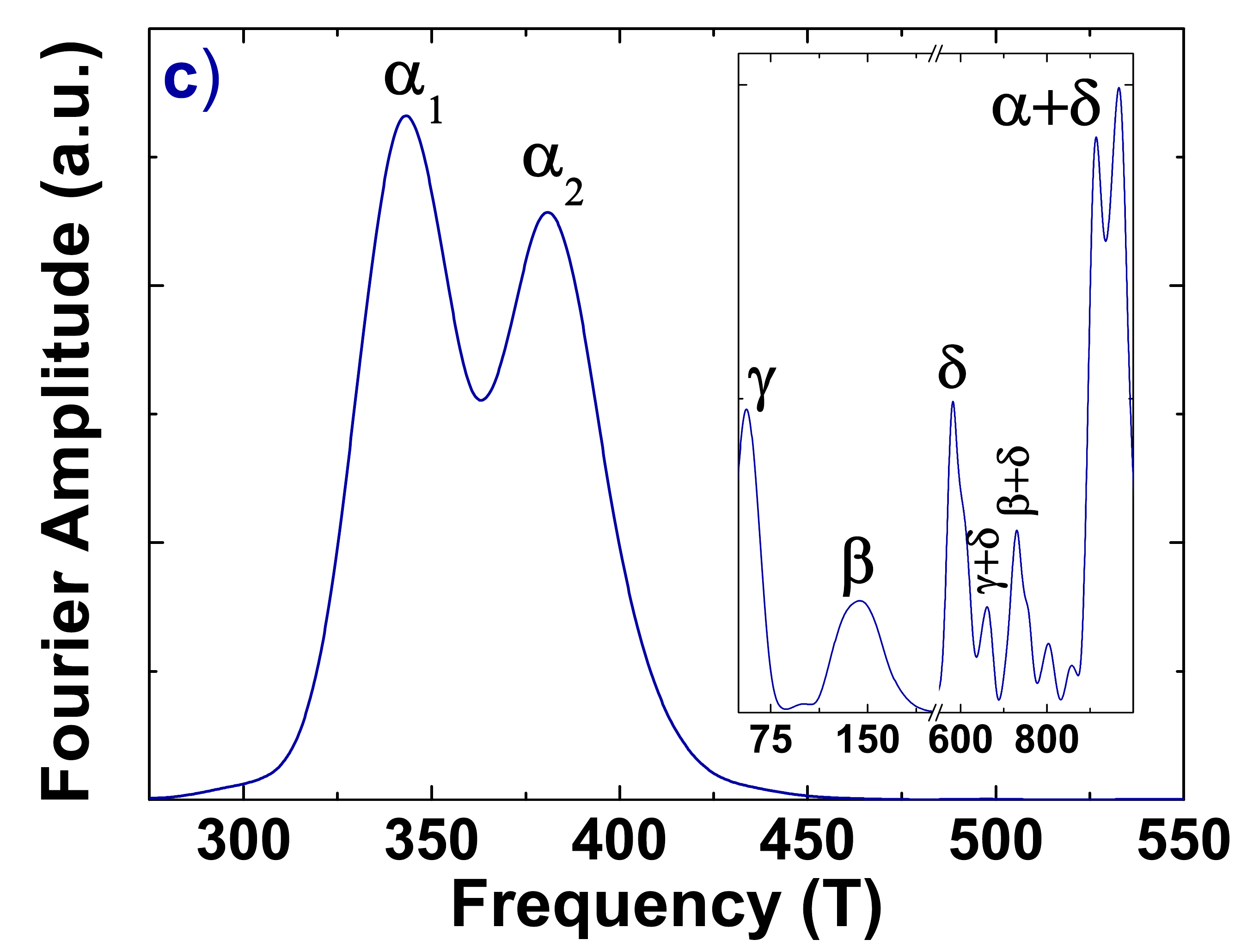} 
 \end{minipage}&
 \begin{minipage}{0.5\textwidth}
\hspace*{-14.5cm}
 \includegraphics[width=0.5\textwidth,bb=0 0 800 600]{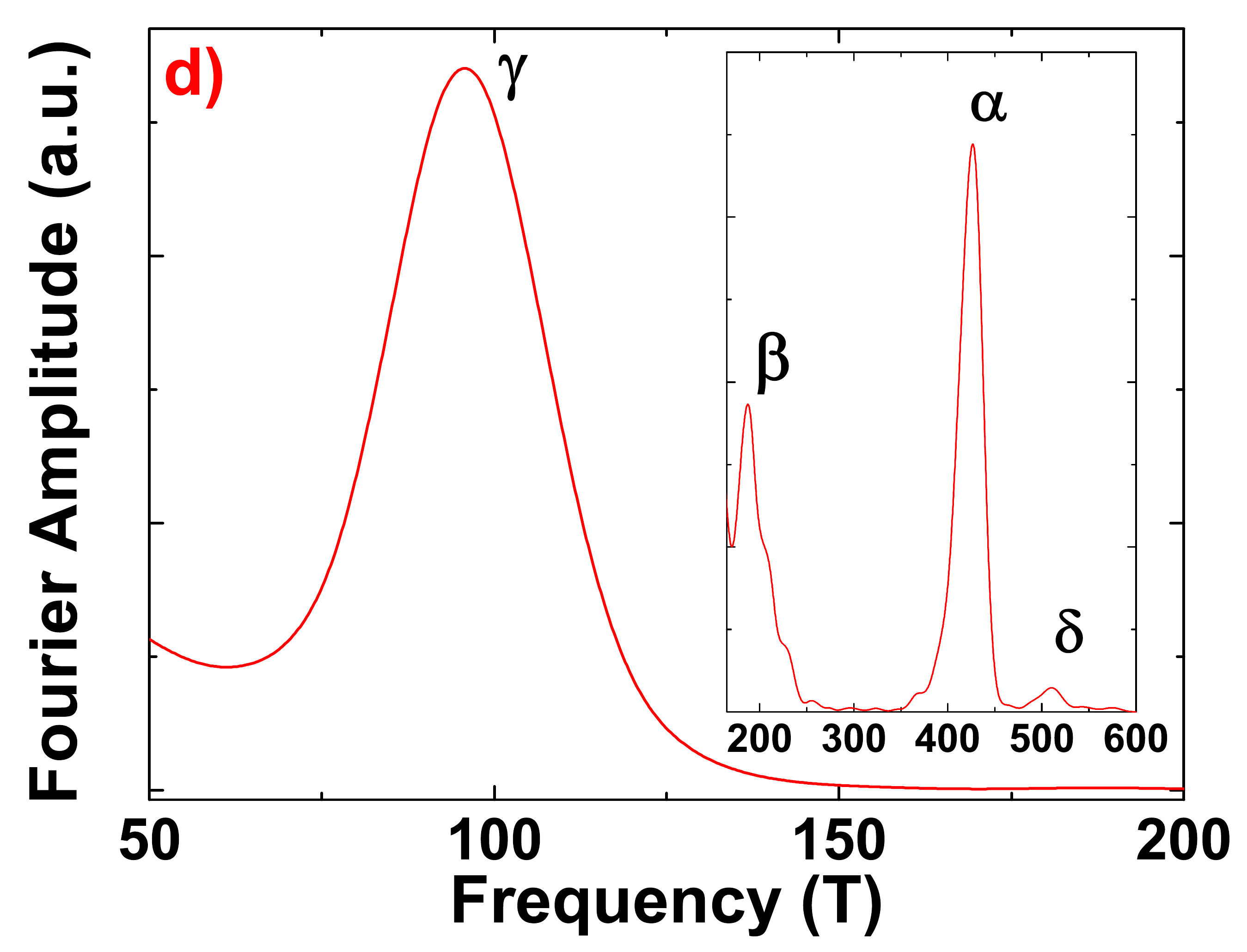} 
 \end{minipage}
 \end{tabular}
\caption{Background subtracted
$\rho_{ab}$($T$) vs. $T$ for a) EuFe$_{2}$As$_{2}$ and b) BaFe$_{2}$As$_{2}$ single crystals at $0^{\circ}$ and $25^{\circ}$. Here, 0$^{\circ}$ means $H$ parallel to the sample surface (ab-plane).
c-d) Fast Fourier Transform (FFT) of the oscillations in Fig. 1a, respectively.}
\end{figure}

Except for $F_{\delta}$, all other pockets are slightly larger in Ba122 as compared to Eu122, suggesting a subtle change of fermiology.  In addition, three magnetic breakdown orbits are observed at higher frequencies in Eu122 corresponding to combinations of the previous pockets ($\gamma+\delta$, $\beta+\delta$, and $\alpha+\delta$). These orbits are due to a magnetic-field-induced  tunneling from one part of the FS to another and it is possible that similar frequencies also exist for the other 122 compounds, but were not detected due to sample quality. Interestingly, Eu122 presents a 
band splitting in the predominant $\alpha$ pocket, in agreement with ARPES measurements below $T_{SDW}$, indicating a small corrugation of the FS.

The next step for a better understanting of such differences is to study the temperature dependence of the observed frequencies. Fig. 3a shows the  Fourier content with increasing temperatures for the Eu122 compound. The corresponding suppression of the SdH amplitudes as a function of the temperature is shown in Fig. 3b and 3c for Eu122 and Ba122, respectively. Such suppression allows us to extract the effective masses by
fitting the temperature dependence of the oscillation amplitude with the thermal damping term
$R_{T}=X/\mathrm{sinh}(X)$ of the Lifshitz-Kosevick (LK) formalism, where $X = 14.69m^{*}T/B$
and $m^{*}$ is the effective mass and $1/B$ is the average inverse field of the Fourier window.
The obtained effective masses are shown in the Fig. 3b. A comparison
with Ba122 effective masses (see Table I) reveal that $m^{*}$ is systematically larger for
Eu122, indicating a slightly higher degree of correlation in the latter. One the other hand, the estimated
Dingle temperature ($T_{D}$)  is roughly the same for all pockets of both compounds. 

\begin{figure}[!ht]
\includegraphics[width=0.35\textwidth,bb=0 0 800 600]{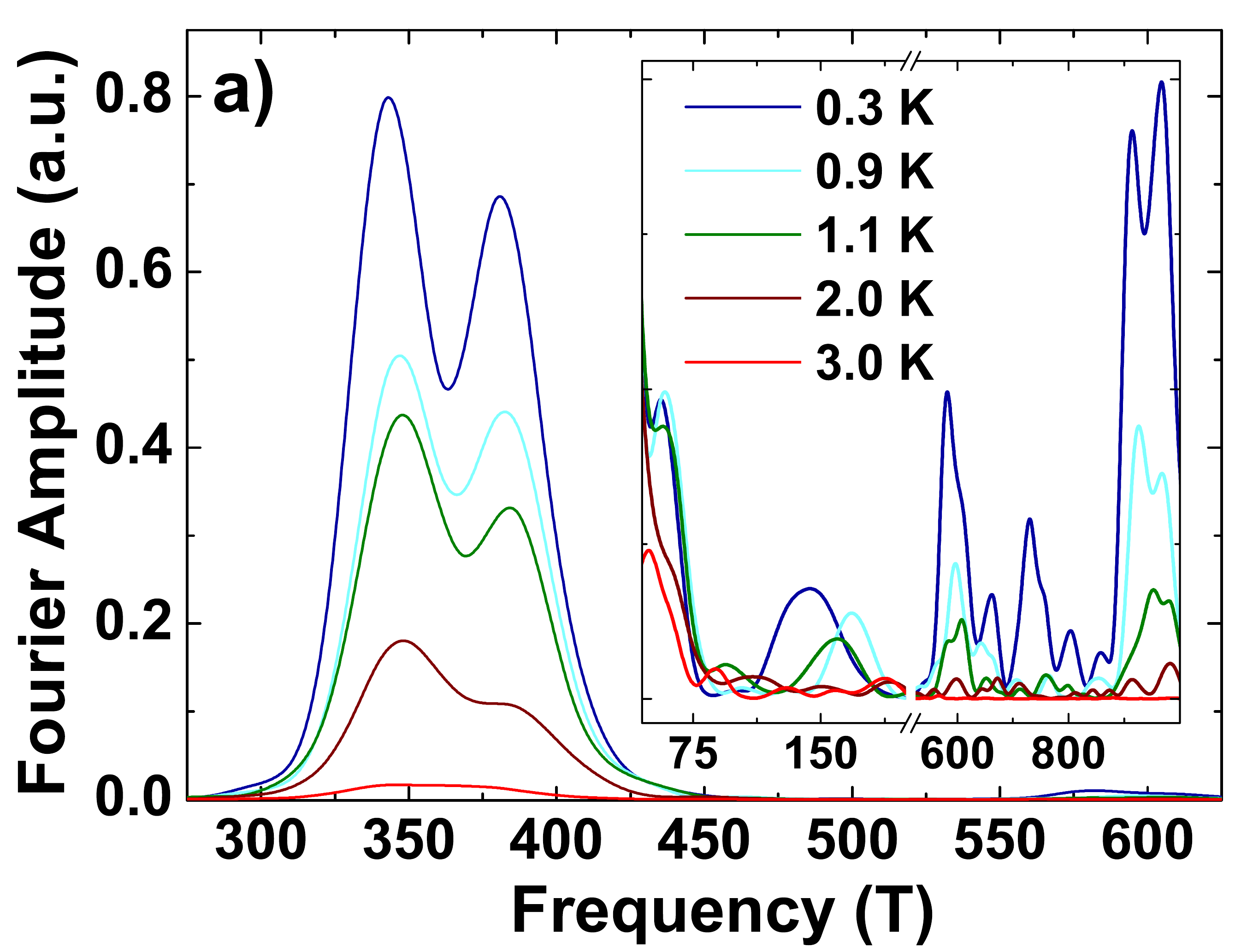}\\
\includegraphics[width=0.35\textwidth,bb=0 0 800 600]{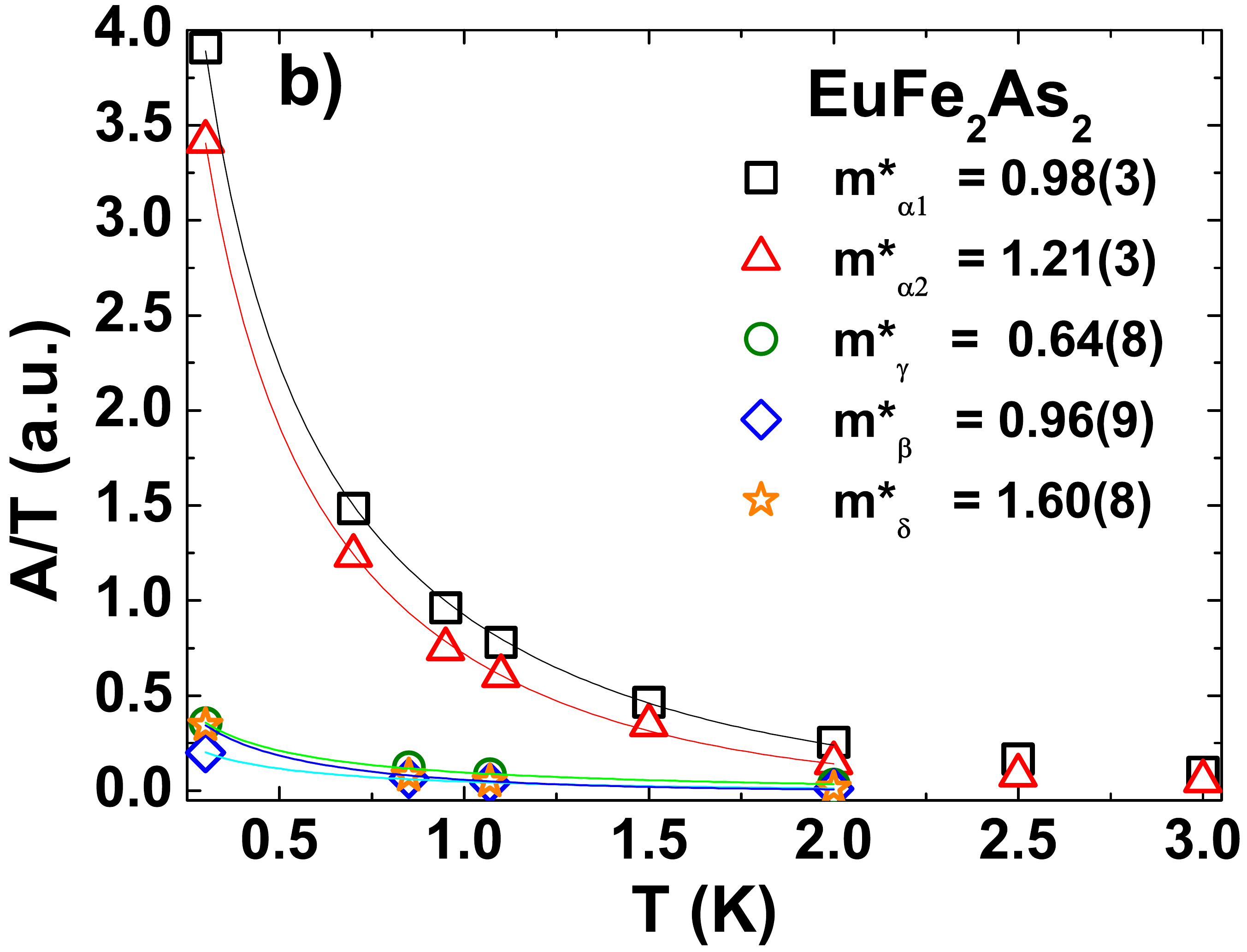}\\
\includegraphics[width=0.35\textwidth,bb=0 0 800 600]{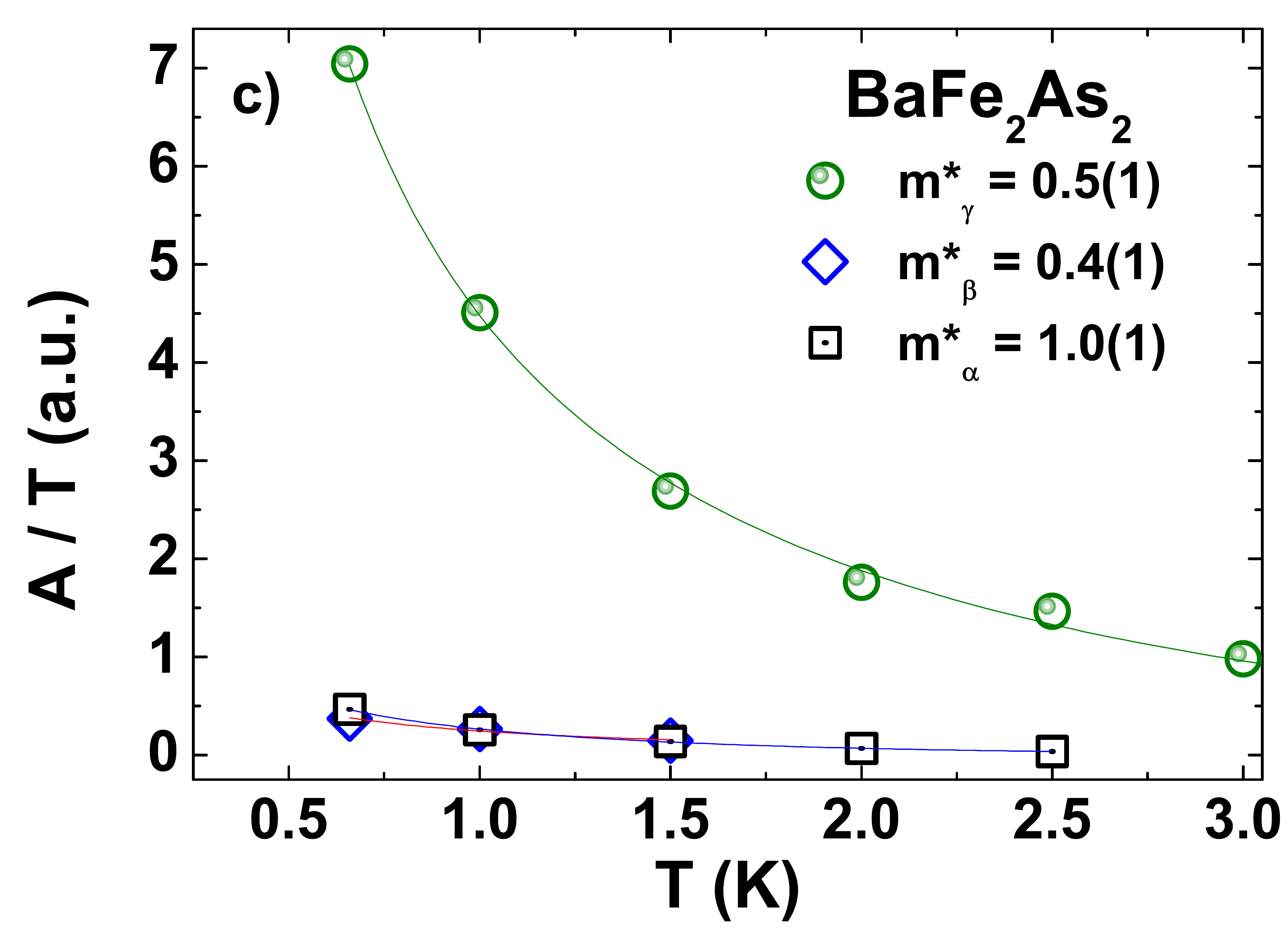}
 \caption{a) FFT amplitude vs frequency for several
temperatures. The inset displays the less intense peaks. b) Fourier amplitude as a function of temperature for the observed pockets in Eu122 and c) in Ba122.}
\end{figure}

\begin{table*}
\begin{centering}
\begin{tabular}{|c||c|c|c|c|c|c|c|c|}
\hline 
  &  \multicolumn{4}{ c|}{BaFe$_{2}$As$_{2}$}  & \multicolumn{4}{ c|}{ EuFe$_{2}$As$_{2}$}   \tabularnewline
\hline
Pocket  & $F (T)$  & $A/A_{BZ}$ ($\%$) & $m^{*}/m^{*}_{e}$ & $T_{D}$ (K) & F (T)  & A/A$_{BZ}$  & $m^{*}/m^{*}_{e}$ &  $T_{D}$  \tabularnewline
\hline 
$\gamma$   &   90(10)   &  0.3  & 0.5(1)  & 5(2)   & 60(10)                    &    0.2        &    0.64(8)               &   -      \tabularnewline
$\beta$       &  190(10)  &  0.6  &  0.4(1)  & 4(1)   & 150(10)                  &    0.6        &    0.96(8)               &   4(1)  \tabularnewline
$\alpha$     &  430(10)  &  1.4  &  1.0(1)  & 4(1)  & 340(10)/380(10)     &    1.3       &    0.98(3)/1.21(3)  &    4(1)  \tabularnewline
$\delta$      & 510(10)   &  1.7  &    -        &  3(1)   & 580(10)                  &      1.4     &    1.60(8)              &    3(1)  \tabularnewline
\hline 
\end{tabular}
\par\end{centering}

\label{Tab_AG} \caption{Comparison between SdH experimental
 values of Ba122 and Eu122.}
\end{table*}

In order to further investigate the differences between the studied compounds, we now turn our attention to the angle dependence of the extremal orbits, shown in Fig. 4. Interestingly, the anisotropy of the pockets evidences the most contrasting behavior of our study. In one hand, Fig. 4a presents a fairly isotropic behavior for the low-frequency $\gamma$ (hole) and $\beta$ (electron) pockets in Eu122. Such isotropy extends over the entire angular range and suggests the presence of largely three-dimensional isotropic FS sections. On the other hand, Fig. 4b shows the anisotropic behavior of $\gamma$ and $\beta$-pockets in Ba122. These branches are associated with large quasi-2D cylinders that show the expected $1/cos(\theta)$ dependence (solid lines) and ellipsity of $\sim 5$. 
 On the other hand, the $\alpha$ (hole) pocket for both compounds is found to present a weaker angular dependence that cannot be fitted to $1/cos(\theta)$. Such behavior suggests that this more isotropic and $3$D particular pocket remains unaltered in the series. 
Finally, we note that the $\delta$ pocket in Eu122
 appears in a narrower angular range. One possible reason for that behavior is the signal-to-noise ratio due to  the very weak FFT peak amplitude. Another possibility could be an anisotropic g-factor, in which case the spin splitting term plays a role in suppressing the amplitude of the oscillations throughout a certain angular range.

\begin{figure}[!ht]
\includegraphics[width=0.4\textwidth,bb=0 0 800 600]{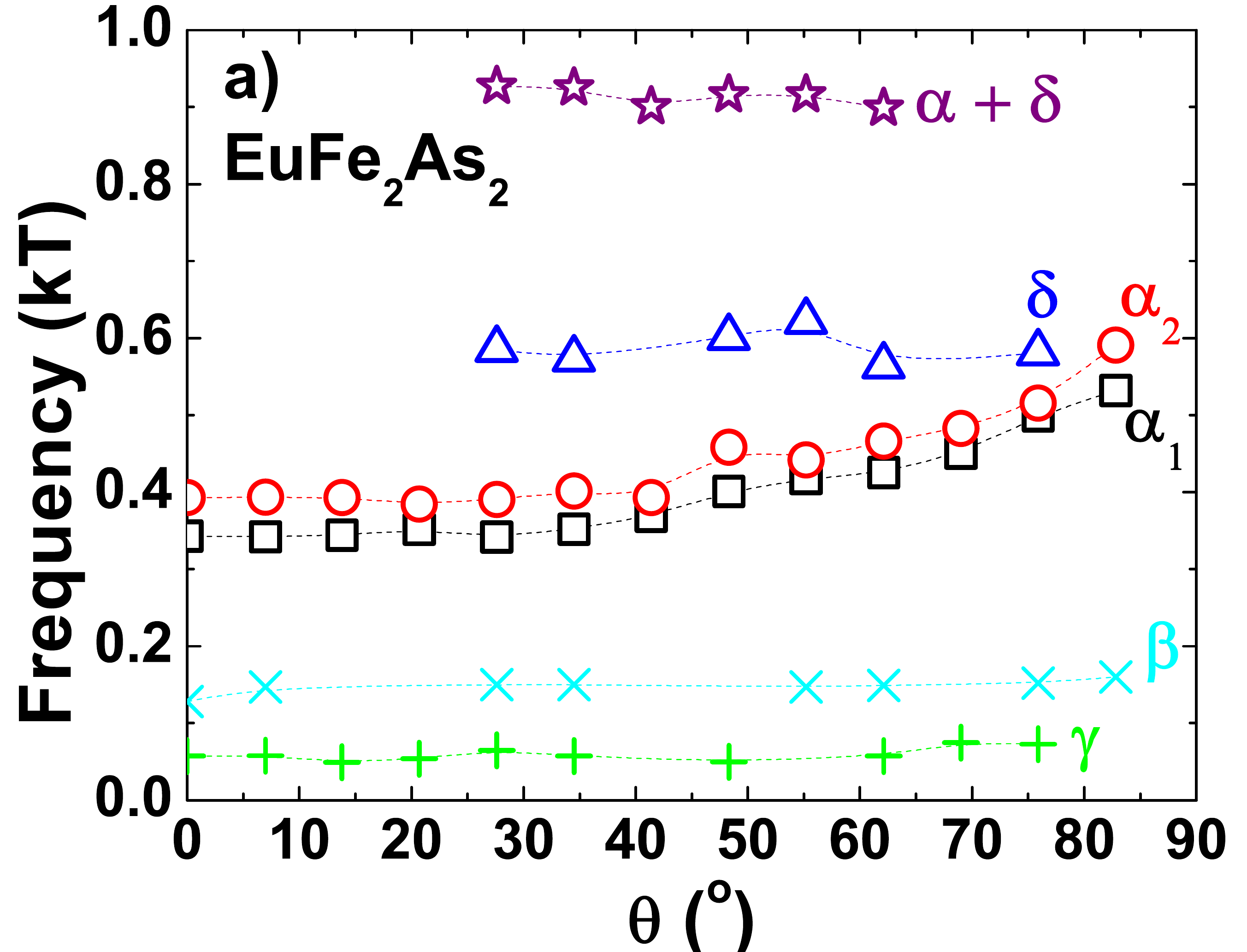}\\
\includegraphics[width=0.4\textwidth,bb=0 0 800 600]{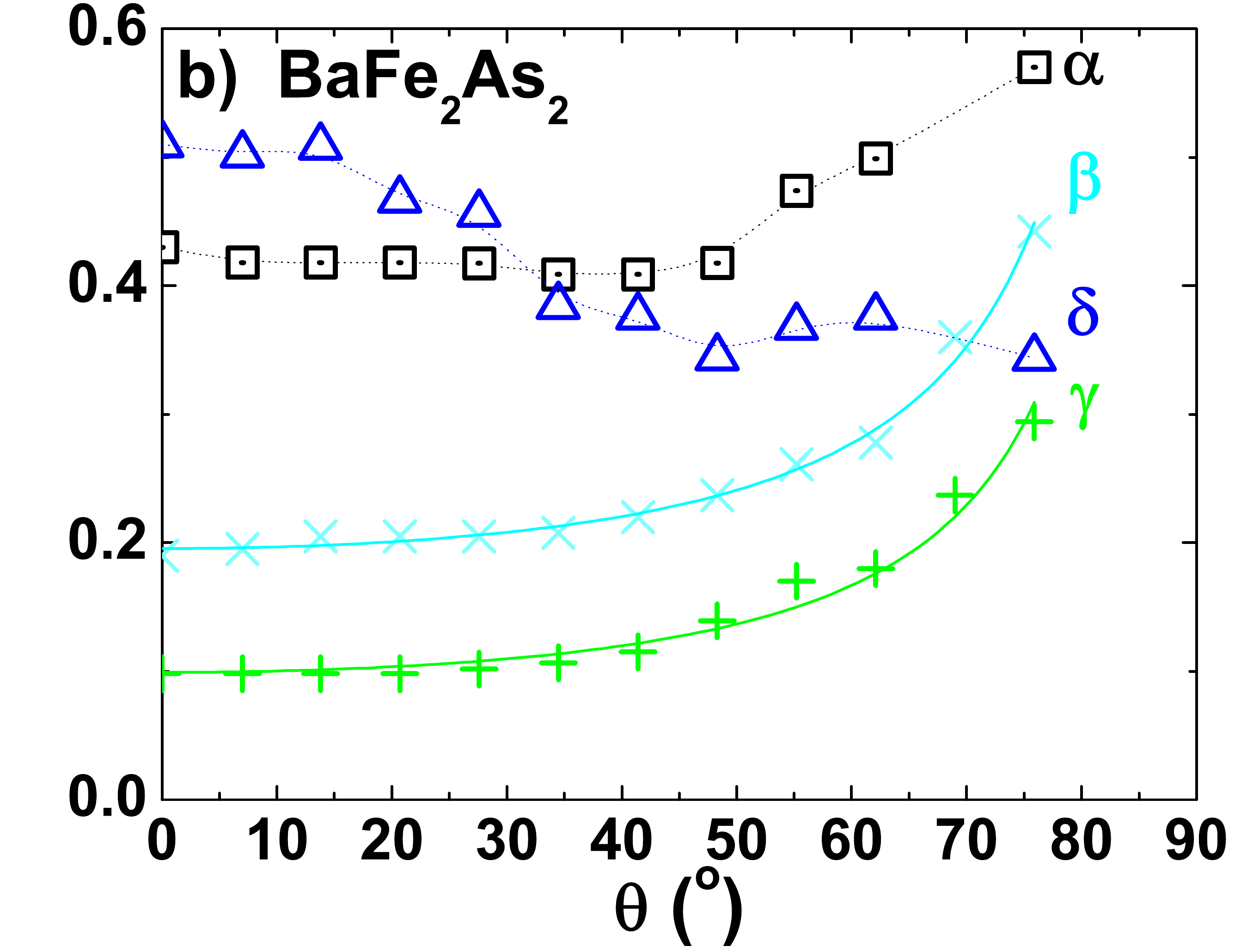}
 \caption{Field orientation dependence of the observed QO frequencies.}
\end{figure}

All the above allow us to conclude that the cylindrical surfaces in Ba122 are much more 2D than in Eu122, suggesting that the planar ($xy/x^{2}-y^{2}$) orbital contribution to the Fe $3d$ bands
is higher in the Ba-member. One could expect that lower dimensionality
 would favor a higher $T_{SDW}$ in Ba122, as compared to Eu122. However, the opposite behavior is verified experimentally with $T_{SDW}$ being $50$ K lower in
Ba122. As the SDW state in these materials is believed to be associated with
the itinerant Fe $3d$ bands, an increase in the planar character of the bands could destabilize
the $3$D itinerant SDW state. It is worth mentioning that our macroscopic understanding of these materials is in agreement with microscopic spin probe techniques, such as Electron Spin Resonance (ESR). Recent ESR measurements also revealed an increase in the anisotropy and localization of the Fe $3d$ bands at the FeAs plane in the
Ba-rich extreme of the series Ba$_{1-x}$Eu$_{x}$Fe$_{2}$As$_{2}$ \cite{Rosa, SS}. Moreover, ESR data also pointed to a similar density of states for both end-compounds. Thus, the higher effective masses observed for Eu122 in the present report cannot be simply attributed to different band widths. 
 It is important to note that such increase in the correlations could also help the increase of $T_{SDW}$ in Eu122. However, there is no systematic increase in the Stoner factor obtained from the magnetic susceptibility along the series, indicating that a small change in the correlations do not play the dominant role in these compounds \cite{Rosa}.

Enlightened by microscopic measurements and calculations, we comment on the possible role of local distortions in determining these differences in the FS properties. Previous EXAFS
(Extended X-Ray Absorption Fine Structure) measurements
 on In-grown Ba122 single crystals revealed that, as the SDW is suppressed with chemical substitution (K and Co) or applied pressure, the Fe-As distance of the pure compound also was suppressed \cite{Granado}. In addition, DFT+DFMT also predicts an effective mass enhancement and an decrease in the planar $xy$ 
orbital contribution as one increases the Fe-As 
distance \cite{KH}. In this context, our data suggest that the As is further away from the
FeAs plane, causing an increase in the correlations. In fact, the As distance from the FeAs plane ($z_{As}$) is known to be higher in Eu122 ($z_{As} = 0.362$, \cite{Raffius}) than in Ba122
($z_{As} = 0.3545(1)$, \cite{Rotter}). Further EXAFS measurements will help us to confirm such scenario.

In summary, we have measured quantum oscillations in the electrical resistivity of EuFe$_{2}$As$_{2}$ and BaFe$_{2}$As$_{2}$ high quality single crystals. Five low frequency pockets are observed with subtle differences between the studied compounds.
Our results show that the effective masses for Eu122 are systematically larger than in the Ba122 member, indicating that the former is a more correlated metal. Moreover, the observed pockets in Ba122 are more anisotropic and two-dimensional, suggesting an orbital differentiation of the Fe $3d$ orbitals
with increasing planar orbital contribution. Our findings may shed new light on the understanding
of the spin-density wave phase suppression toward a superconducting state.

\section{\label{sec:ack} Acknowledgments}

This work was supported by FAPESP-SP, AFOSR MURI, CNPq and FINEP-Brazil. L.B. is supported by DOE-BES through award DE-SC0002613.
The NHMFL is supported by NSF through NSF-DMR-0084173 and the
State of Florida.

\end{document}